\documentclass{amsart}
\usepackage{amsmath,amsfonts}
\usepackage{amscd,amsthm}
\usepackage{graphicx}

\theoremstyle{plain} 

\newtheorem{theorem}{Theorem}[section]
\newtheorem{corollary}[theorem]{Corollary}

\newtheorem{proposition}[theorem]{Proposition}

\newtheorem{remark}[theorem]{Remark}

\newcommand{\documentdate}{12 December 2010}

\begin{document}

\begin{titlepage}

\includegraphics[height=3.5cm]{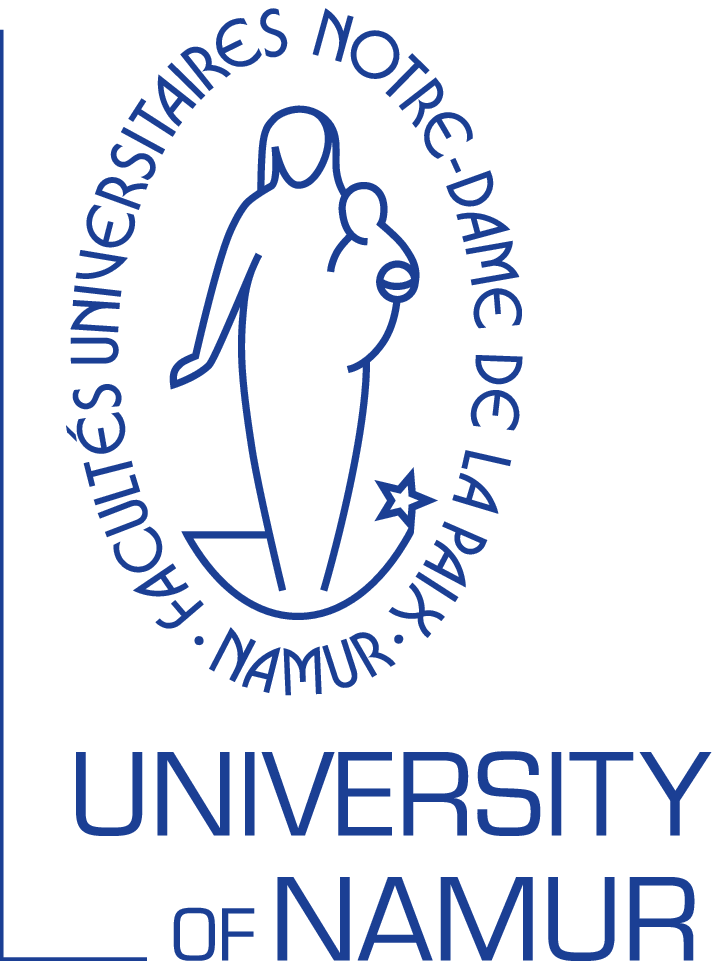}

\vspace*{1cm}
\hspace*{1cm}
\fbox{\rule[-3cm]{0cm}{6cm}\begin{minipage}[c]{12cm}
\begin{center}
{\Large High order explicit symplectic integrators\\ for the Discrete Non Linear
  Schr\"odinger equation}\\
\mbox{}\\
by Jehan Boreux, Timoteo Carletti and Charles Hubaux\\
\mbox{}\\
Report naXys-09-2010 \hspace*{20mm} \documentdate 
\end{center}
\end{minipage}
}

\vspace{2cm}
\begin{center}
\includegraphics[height=3.5cm]{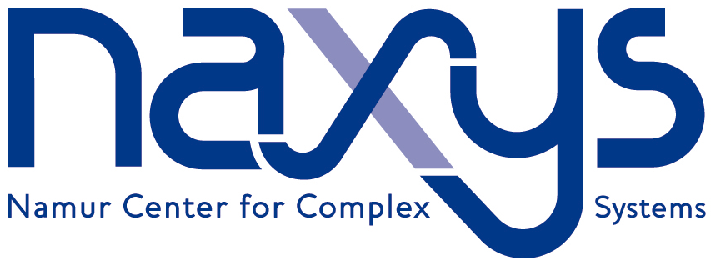}

\vspace{1.5cm}
{\Large \bf Namur Center for Complex Systems}

{\large
University of Namur\\
8, rempart de la vierge, B5000 Namur (Belgium)\\*[2ex]
{\tt http://www.naxys.be}}

\end{center}

\end{titlepage}

\newpage

\title[High order explicit symplectic integrators for the DNLS]{High order explicit symplectic integrators for the Discrete Non Linear
  Schr\"odinger equation.}

\author{Jehan Boreux}
\address{naXys Namur Center for Complex Systems, University of Namur 
\\8, Rempart de la Vierge, 5000 Namur, Belgium.\\
jehan.boreux@fundp.ac.be.}

\author{Timoteo Carletti}
\address{naXys Namur Center for Complex Systems, University of Namur 
\\8, Rempart de la Vierge, 5000 Namur, Belgium.\\
timoteo.carletti@fundp.ac.be.}

\author{Charles Hubaux}
\address{naXys Namur Center for Complex Systems, University of Namur 
\\8, Rempart de la Vierge, 5000 Namur, Belgium.\\
charles.hubaux@fundp.ac.be.}

\maketitle

\begin{abstract}
We propose a family of reliable symplectic integrators
adapted to the Discrete Non--Linear Schr\"odinger equation; based on an idea
of Yoshida~\cite{Yo1990} we can construct high order numerical schemes, that
result to be explicit methods and thus very fast. The
performances of the integrators are discussed, studied as functions of the
integration time step and compared with some non symplectic methods.
\end{abstract}

\maketitle
\section{Introduction}
\label{sec:intro}
The one dimensional Nonlinear Schr\"odinger System (NLS)~\cite{K2009} : 
\begin{equation}
  \label{eq:nls}
\begin{cases}
 \;\; i \partial_t q+\partial_{xx}q+2q^2p=0\\
-i \partial_t  p+\partial_{xx}p+2p^2q=0\, ,
\end{cases}
\end{equation}
under periodic boundary conditions, $\left(p(x + L, t), q(x + L, t)\right) =
\left(p(x, t), q(x, t)\right)$ for all $x$ and $t$, is a widely studied
multiparticles system subjected to nonlinear interactions, that can be used to
model several relevant physical phenomena~\cite{Porter2009}, ranging from
optics to solid state and atomic physics, for instance Bose-Einstein
condensates. 

Often scientists, once modeling physical nonlinear phenomena, impose the
 condition~\footnote{Where a $+$ sign represents the defocusing case hereby
  considered, a $-$ sign represents the focusing one and where $\overline{a}$
  denotes the   
complex conjugate of the complex number $a$.} $\overline{p(t)} = \pm q(t)$ for
all 
$t$, 
 in this case the system reduces
to the standard cubic NLS equation. We hereby adopt the 
viewpoint of~\cite{KS2001} where the conjugacy relation between $p(t)$ and
$q(t)$ is not imposed a priori. We will nevertheless show that if this relation
holds at $t=0$ then it will be preserved by our numerical integration scheme.

Most properties of NLS are related to the asymptotic behavior of its
solutions, there is thus a strong need for integration schemes,
allowing large step sizes, to easily cover large time spans, without
degrading the efficiency of the numerical method hence avoiding the
introduction 
of spurious outcome in the numerical simulations. These goals can be achieved
using symplectic 
integrators, that are specifically designed to preserve the energy and
possibly other first integral of the system. Our analysis will be performed on
the following one dimensional discretization of the Eq.~\eqref{eq:nls}, DNLS
for short~\cite{KS2001}:
\begin{equation}\forall l\in\{1,\dots,N\}\quad 
  \label{eq:dnls}
  \begin{cases}
    \dot{p_l}&=-\frac{i}{h^2}\left(p_{l+1}-2p_l+p_{l-1}\right)-2ip_l^2q_l\\
    \dot{q_l}&=\frac{i}{h^2}\left(q_{l+1}-2q_l+q_{l-1}\right)+2ip_lq_l^2\, ,
  \end{cases}
\end{equation}
using the periodic boundary conditions $q_{l+N}=q_l$ and $p_{l+N}=p_l$ for all
$l\in\{1,\dots,N\}$. Let us observe that the solutions of the DNLS are
$\mathcal{O}(h^2)$ 
close to the solutions of the PDE~\eqref{eq:nls}, thus in the limit
$h\rightarrow 0$, the former converge to the solutions of the latter.

The integration method hereby proposed is based on the development of an
idea introduced by Yoshida~\cite{Yo1990}. The strong improvement allowed by
this method with respect to other ones available in the literature, relies
on the fact that we are able to provide {\em symplectic} integrators,
that are {\em explicit} ones, thus very fast, and whose {\em order} can be
easily as high as the eighth one. Other methods based on
the construction of generating functions, see for
instance~\cite{CS1990,MCLA1992} are not particularly suitable because the
Hamiltonian function describing the DNLS cannot be split in the form
$H(p,q)=T(p)+V(q)$, namely it is not a potential Hamiltonian system. This
implies in fact that the obtained numerical integration schemes are implicit
methods, and thus a large amount of CPU time is used to 
compute the new position, or the new momentum, using some Newton--like
method. Let us observe that Gauss' methods~\cite{HLW2006}, namely
symplectic versions of Runge--Kutta, are also implicit ones and thus they
suffer of the above mentioned limitations.

Finally let us observe that, because the system cannot be put in the form of
a perturbation of an integrable system, the method proposed in~\cite{LR2001}
is no more useful here: one can easily get a second order method and
with some additional work also a fourth order method~\cite{SKKF2009}, but no
higher orders are obtainable. 

The paper is organized as follows. In the next section we briefly recall the
Hamiltonian structure of the non linear Schr\"odinger equation on a one
dimensional lattice;
then Section~\ref{sec:symplsch} will be devoted to the presentation and to the
construction of the symplectic integrator, whose properties will be
numerically studied in Section~\ref{sec:results}. Finally we sum up and draw
our conclusions in Section~\ref{sec:concl}.

\section{The non linear Schr\"odinger equation on a lattice}
\label{sec:nlsd}

The system~\eqref{eq:nls} will be studied imposing the standard spatial
discretization, see for instance~\cite{KS2001}, thereby called {\em Diagonal} 
NLS, that reads:
\begin{equation}\forall l\in\{1,\dots,N\}\quad 
  \label{eq:maineq}
  \begin{cases}
    \dot{p_l}&=-\frac{i}{h^2}\left(p_{l+1}-2p_l+p_{l-1}\right)-2ip_l^2q_l\\
    \dot{q_l}&=\frac{i}{h^2}\left(q_{l+1}-2q_l+q_{l-1}\right)+2ip_lq_l^2\, ,
  \end{cases}
\end{equation}
using the periodic boundary conditions:
\begin{equation}
  \label{eq:perbc}
  q_{l+N}=q_l \quad \text{and}\quad
  p_{l+N}=p_l\quad \forall l\in\{1,\dots,N\}\, .
\end{equation}
Observe that $h$ plays the role of spatial discretization parameter and
thus the first terms on the right hand sides stem from the discretized second
order spatial derivatives.

Let us remark that one could be interested in studying directly the
system~\eqref{eq:maineq} as a relevant model of non--linear interaction on a
discrete one dimensional of lattice.

One can easily prove that the DNLS can be cast into the Hamiltonian
formalism using the following Hamilton function
\begin{equation}
  \label{eq:hamilt}
  H(\vec{p},\vec{q})=-i\sum_{l=1}^N\left[\frac{(p_{l+1}-p_l)(q_{l+1}-q_l)}{h^2}-p_l^2q_l^2\right]\,
  ,
\end{equation}
and moreover the variables $(p_l,q_l)\in\mathbb{C}^n$ do satisfy the standard
Poisson equations 
\begin{equation*}
  \{p_l,p_m\}=\{q_l,q_m\}=0 \quad \text{and}\quad   \{p_l,q_m\}=\delta_{l,m}
  \quad \forall l,m\in\{1,\dots,N\}\, .
\end{equation*}

\begin{remark}
  Let us observe that the DNLS system~\eqref{eq:maineq} possesses another
  first integral independent of the energy, namely:
  \begin{equation}
    \label{eq:secfisrtint}
    I(\vec{p},\vec{q})=\sum_{l=1}^Np_lq_l\, ,
  \end{equation}
that under the assumption $\overline{p(t)}=q(t)$, is usually called the
{\em mass} of the system.
\end{remark}

The aim of this work is to define a family of symplectic integrators based on
the Yoshida symplectic scheme~\cite{Yo1990} adapted to the
DNLS system~\eqref{eq:maineq} and to study their numerical properties in terms
of preserved quantities, CPU time needed and accuracy of the results.

Let us observe that system~\eqref{eq:maineq} is not quasi--integrable, i.e. it
cannot be decomposed as the sum of an integrable one and a \lq\lq small
perturbation\rq\rq , thus we cannot use the high order symplectic schemes
$SABA_n$ or $SBAB_n$ proposed by Laskar and Robutel~\cite{LR2001}, whose
accuracy strongly relies on the smallness of the perturbation.

\section{A symplectic scheme}
\label{sec:symplsch}

For a sake of completeness, let us briefly recall the integration method
proposed in~\cite{Yo1990} to numerically reconstruct the orbit of an
Hamiltonian systems $H(\vec{p},\vec{q})$.  

Given any initial condition 
$\left(\vec{p}(0),\vec{q}(0)\right)$ and an integration time span $[0,T]$, we
proceed by decomposing the integration interval into small pieces of fixed
size $\tau$. The method results thus in a fixed step size integration
method. Then, in each small interval, we approximate the time $\tau$ flow of the
Hamiltonian systems 
\begin{equation*}
  \begin{cases}
    \dot{\vec{p}}&=-\frac{\partial H}{\partial \vec{q}}\\
    \dot{\vec{q}}&=\frac{\partial H}{\partial \vec{p}}\, ,
  \end{cases}
\end{equation*}
 by a 
composition of {\em basic symplectic flows} of the form $\exp\left(\tau c_jL_A\right)$
and $\exp\left(\tau d_jL_B\right)$, where $A(\vec{p},\vec{q})$ and $B(\vec{p},\vec{q})$
are two suitable functions providing a decomposition of
the Hamilton function, i.e. $H=A+B$, $(c_j,d_j)$ suitable constants to
achieve the wanted order of precision of the integration scheme and $\exp\left(\tau
L_H\right)$ is a shorthand notation to denote the flow at time $\tau$ of the Hamilton
function $H$. More precisely we are looking for
\begin{equation}
  \label{eq:approxflow}
  \exp\left(\tau L_{A+B}\right)=  \exp\left(\tau c_1L_{A}\right)\circ \exp\left(\tau d_1L_{B}\right) \circ
  \dots \circ \exp\left(\tau c_kL_{A}\right)\circ \exp\left(\tau
  d_kL_{B}\right)+\mathcal{O}(\tau^n)\, , 
\end{equation}
for some positive integers $k$ and $n$.

The method is
particularly efficient once the maps $\exp\left(\tau c_jL_A\right)$ and $\exp\left(\tau
d_jL_B\right)$ are explicitely computable, which requires $A$ and $B$ to be {\em
  simple enough}. In particular this is true whenever $A$ and $B$ depend only
upon one group of canonical variables and thus the Hamiltonian is of the
so-called potential
form. Of course this is a strong requirement that cannot always be achieved by
suitable change of coordinates. In the following we propose the decomposition of
the Hamilton function~\eqref{eq:hamilt} given by
\begin{equation}
  \label{eq:splitAB}
A(\vec{p},\vec{q})=i\sum_{l=1}^Np_l^2q_l^2 \quad \text{and}\quad
B(\vec{p},\vec{q})=-i\sum_{l=1}^N\frac{(p_{l+1}-p_l)(q_{l+1}-q_l)}{h^2} \, .
\end{equation}
Let us observe that even if $A$ and $B$ both depend on all the canonical
variables, we are able to explicitely compute $\exp\left(\tau L_A\right)$
and $\exp\left(\tau L_B\right)$ (see \S~\ref{ssec:expA} and
\S~\ref{ssec:expB}) and thus to propose a {\em completely explicit} method.

\subsection{Computation of $\exp\left(\tau L_A\right)$}
\label{ssec:expA}

The equations of motion of the system with \lq\lq Hamilton function\rq\rq $A$
are given by:
\begin{equation}
  \label{eq:expA1}\forall l\in\{1,\dots,N\}\quad 
  \begin{cases}
    \dot{p}_l&=-2ip_l^2q_l\\
    \dot{q}_l&=2ip_lq^2_l\, ,
  \end{cases}
\end{equation}
from which it trivially follows that $p_lq_l=C_l$ is a first integral for all
$l$, in fact: 
\begin{equation*}
  \frac{d}{dt}C_l\rvert_{\text{flow~\eqref{eq:expA1}}}=\dot{p}_lq_l+p_l\dot{q}_l=-2ip_l^2q^2_l+2ip_l^2q^2_l=0\, .
\end{equation*}
Hence~\eqref{eq:expA1} simplifies into
\begin{equation}
  \label{eq:expA2}\forall l\in\{1,\dots,N\}\quad 
  \begin{cases}
    \dot{p}_l&=-2ip_lC_l\\
    \dot{q}_l&=2iq_lC_l\, ,
  \end{cases}
\end{equation}
whose solution with initial datum $(p_l(0),q_l(0))_{l=1,\dots , N}$ is for all $t$:
\begin{equation}
  \label{eq:expA3}\forall l\in\{1,\dots,N\}\quad 
  \begin{cases}
    p_l(t)&=e^{-2iC_lt}p_l(0)\\
    q_l(t)&=e^{2iC_lt}q_l(0)\, .
  \end{cases}
\end{equation}
Finally the time $\tau$ flow of $A$ is given by:
\begin{eqnarray}
  \label{eq:expA4}
(\vec{p^{\prime}},\vec{q^{\prime}})^{\rm T}&=&e^{\tau L_A}(\vec{p},\vec{q})^{\rm T}=e^{\tau
  L_A}(p_1,\dots,p_N,q_1,\dots,q_N)^{\rm T}\notag\\ &=&
(e^{-2ip_1q_1\tau}p_1,\dots,e^{-2ip_Nq_N\tau}p_N,e^{2ip_1q_1\tau}q_1,\dots,e^{2ip_Nq_N\tau}q_N)^{\rm T}\,
,
\end{eqnarray}
where $^{\rm T}$ denotes the transposed of the vector.

\subsection{Computation of $\exp\left(\tau L_B\right)$}
\label{ssec:expB}

The equations of motion corresponding to the $B$ part of the Hamiltonian
function are:
\begin{equation}\forall l\in\{1,\dots,N\}\quad 
  \label{eq:expB1}
  \begin{cases}
    \dot{p_l}&=-\frac{i}{h^2}\left(p_{l+1}-2p_l+p_{l-1}\right)\\
    \dot{q_l}&=\frac{i}{h^2}\left(q_{l+1}-2q_l+q_{l-1}\right)\, ,
  \end{cases}
\end{equation}
that can be cast in a compact form using the periodic boundary
conditions~\eqref{eq:perbc} by introducing the {\em circulant matrix}~\cite{Davis1979}
$M$ obtained by cyclically permute the $N$--vector $(-2,1,0,\dots,0,1)$:
\begin{equation}
  \label{eq:matrixM}
  M=\left(
    \begin{matrix}
      -2 & 1 & 0 & \hdots & \hdots & 0 & 1\\
       1 & -2 & 1 & 0 & \hdots & \hdots & 0 \\
       0 & 1 & -2 & 1 & 0 & \hdots & 0 \\
       \vdots & \ddots & \ddots & \ddots & \ddots &  & \vdots \\
       \vdots &  & \ddots & \ddots & \ddots &  & \vdots \\
       \vdots &  &  & \ddots & \ddots & \ddots & \vdots \\
       1 & 0 & \hdots &\hdots & 0 & 1 & -2
    \end{matrix}
\right)\, ,
\end{equation}
that is the linear diagonal system
\begin{equation}
  \label{eq:expB2}
  \begin{cases}
    \dot{\vec{p}}&=-\frac{i}{h^2}M\vec{p}\\
     \dot{\vec{q}}&=\frac{i}{h^2}M\vec{q}\, .
  \end{cases}
\end{equation}

Thus formally the time $\tau$ flow of the Hamilton function $B$ is given by:
\begin{eqnarray}
  \label{eq:expB3}
(\vec{p^{\prime}},\vec{q^{\prime}})^{\rm T}&=&e^{\tau
  L_B}(\vec{p},\vec{q})^{\rm T}=(e^{-\frac{i}{h^2}\tau M}\vec{p},e^{\frac{i}{h^2}\tau
  M}\vec{q})^{\rm T}\, .
\end{eqnarray}

The circulant matrices have some useful properties that allow one to
explicitely compute their eigenvalues, eigenvectors and hence their exponential.

\begin{proposition}[Circulant matrix]
\label{prop:circmatr}
  Let $M$ be the circulant matrix given by~\eqref{eq:matrixM} and let
  $\rho=e^{2\pi i/N}$, then
  \begin{enumerate}
  \item The eigenvalues of the matrix $M$ are given by:
    $\mu_j=-2+\rho^{j}+\rho^{j(N-1)}$, 
    $j=0,\dots,N-1$;
  \item For $j=0,\dots,N-1$, the eigenvector associated to $\mu_j$ is given by
    $$w_j=\frac{1}{\sqrt{N}}\left(1,\rho^j,\rho^{2j},\dots,\rho^{j(N-1)}\right)^{\rm
      T} \, ;$$ 
\item Let $W$ be the matrix whose columns are the eigenvectors
  $w_0,\dots,w_{N-1}$, then $W$ is unitary, namely
  $WW^{\dag}=W^{\dag}W=1_N$. Where $^\dag$ denotes the transposed complex
  conjugated matrix; 
\item The exponential of $M$ is given by:
  \begin{equation*}
    e^M =We^{M_{\text{diag}}}W^{\dag}\Rightarrow \forall
    l,m\in\{1,\dots,N\}\quad (e^M)_{lm}=\frac{1}{N}\sum_{k=1}^N
    e^{\mu_{k-1}}\rho^{(l-m)(k-1)}\, ,
  \end{equation*}
being $M_{\text{diag}}$ the diagonal matrix with the eigenvalues on the diagonal.
  \end{enumerate}
\end{proposition}

\proof
Point (1) can be proved by a direct computation, for all $j\in\{0,\dots,
N-1\}$, one has:
\begin{equation*}
  Mw_j=\frac{1}{\sqrt{N}}M\left(
    \begin{smallmatrix}
      1\\ \rho^j \\ \vdots \\ \rho^{j(N-1)}
    \end{smallmatrix}
\right)=\frac{1}{\sqrt{N}}\left(
    \begin{smallmatrix}
      -2+\rho^j+\rho^{j(N-1)}\\ 1-2\rho^j+\rho^{2j} \\ \vdots \\
      1+\rho^{j(N-2)}-2\rho^{j(N-1)} 
    \end{smallmatrix}
\right)=\frac{1}{\sqrt{N}}\left(
    \begin{smallmatrix}
      -2+\rho^j+\rho^{j(N-1)}\\ \rho^j (\rho^{-j}-2+\rho^{j}) \\ \vdots \\
      \rho^{j(N-1)}(\rho^{-j(N-1)}+\rho^{-j}-2) 
    \end{smallmatrix}
\right)\, ,
\end{equation*}
observing that $\rho^N=1$, we thus have
\begin{equation*}
  Mw_j=\frac{-2+\rho^j+\rho^{j(N-1)}}{\sqrt{N}}\left(
    \begin{smallmatrix}
      1\\ \rho^j \\ \vdots \\
      \rho^{j(N-1)} 
    \end{smallmatrix}
\right)=\mu_j w_j\, .
\end{equation*}

Let us point out that the complex vectors $(w_j)_{j=0,\dots,N-1}$ form an
orthonormal set for 
$\mathbb{C}^N$ (with the standard complex scalar product):
\begin{equation*}
  <w_j,w_k>=\sum_{m=1}^N\overline{w}_{j,m}w_{k,m}=\frac{1}{N}\sum_{m=1}^N\rho^{-j(m-1)}\rho^{k(m-1)}=\frac{1}{N}\sum_{m=1}^N\rho^{-(j-k)(m-1)}=\delta_{jk}\, .
\end{equation*}

Let us also remark that the eigenvalues $\mu_j$ are actually real:
\begin{equation*}
  \mu_j=-2+\rho^j+\rho^{j(N-1)}=-2+\rho^j+\rho^{-j}=-2+2\cos(2\pi j/N)\, ,
\end{equation*}
and moreover the eigenvalues coincide in pairs, more precisely there are
$(N-1)/2$ distinct eigenvalues, if 
$N$ is odd, and $(N-2)/2$ ones if $N$ is even: $\mu_j=\mu_k$
if $(j+k) \mod N=0$.

To prove that $W$ is unitary, let us denote by $X=WW^{\dag}$ and compute the
element $j,k$ of such matrix. By definition 
\begin{equation*}
  X_{jk}=\sum_{l=1}^NW_{jl}W_{lk}^{\dag}=\sum_{l=1}^NW_{jl}\overline{W}_{kl}=\frac{1}{N}\sum_{l=1}^N\rho^{(j-1)(l-1)}\rho^{-(k-1)(l-1)}=\frac{1}{N}\sum_{l=1}^N\rho^{(j-k)(l-1)}=\delta_{jk}\, ,
\end{equation*}
thus $X=1_N$. A similar computation can be done for $W^{\dag}W$.

Finally let us observe that $MW=WM_{\text{diag}}$, in fact one trivially has
\begin{equation*}
  MW=(\mu_0w_0|\hdots|\mu_{N-1}w_{N-1})=WM_{\text{diag}}\, ,
\end{equation*}
and because $W$ is unitary, $W^{-1}=W^{\dag}$, thus
$M=WM_{\text{diag}}W^{\dag}$. This implies that the exponential can be
computed as follows:
\begin{equation*}
  e^{M}=e^{WM_{\text{diag}}W^{\dag}}=We^{M_{\text{diag}}}W^{\dag}\, .
\end{equation*}
The element $l,m$ of $e^M$ is thus given by:
\begin{eqnarray*}
  (e^{M})_{lm}&=&\sum_{k=1}^N W_{lk}e^{\mu_{k-1}}W_{km}^{\dag}=\sum_{k=1}^N
  W_{lk}e^{\mu_{k-1}}\overline{W}_{mk}=\frac{1}{N}\sum_{k=1}^N
  e^{\mu_{k-1}}\rho^{(l-1)(k-1)} \rho^{-(m-1)(k-1)}\\
&=&\frac{1}{N}\sum_{k=1}^N e^{\mu_{k-1}}\rho^{(l-m)(k-1)}\, .
\end{eqnarray*}
\endproof

One can thus explicitely compute the flow of $B$ given by~\eqref{eq:expB3}

\begin{corollary}
  The time $\tau$ flow of the Hamilton function $B$ can be rewritten as:
  \begin{equation}
    \label{eq:expB4}
    \begin{cases}
    \vec{p^{\prime}}&=We^{-\frac{i}{h^2}\tau M_{\text{diag}}}W^{\dag}\vec{p}\\      
    \vec{q^{\prime}}&=We^{\frac{i}{h^2}\tau M_{\text{diag}}}W^{\dag}\vec{q}\, ,
    \end{cases}
  \end{equation}
or explicitely for all $l\in\{1,\dots,N\}$:
\begin{equation}
  \label{eq:expB5}
  \begin{cases}
    p_l^{\prime}=\frac{1}{N}\sum_{m=1}^N\sum_{k=1}^N
    e^{-\frac{i}{h^2}\tau\mu_{k-1}}\rho^{(l-m)(k-1)}p_m\\
q_l^{\prime}=\frac{1}{N}\sum_{m=1}^N\sum_{k=1}^N
e^{\frac{i}{h^2}\tau\mu_{k-1}}\rho^{(l-m)(k-1)}q_m\ .
  \end{cases}
\end{equation}
\end{corollary}

Let us observe that~\eqref{eq:expB5} can be rewritten in compact vector form
in a way inspired by the solution of linear ODE, namely as a linear
combination of 
eigenvectors, as follows:
  \begin{equation}
    \label{eq:expB6}
    \begin{cases}
    \vec{p^{\prime}}&=e^{-\frac{i}{h^2}\tau
      \mu_0}<w_0,\vec{p}(0)>w_0+\dots+e^{-\frac{i}{h^2}\tau
      \mu_{N-1}}<w_{N-1},\vec{p}(0)>w_{N-1}\\
    \vec{q^{\prime}}&=e^{\frac{i}{h^2}\tau
      \mu_0}<w_0,\vec{q}(0)>w_0+\dots+e^{-\frac{i}{h^2}\tau
      \mu_{N-1}}<w_{N-1},\vec{q}(0)>w_{N-1}\, .
    \end{cases}
  \end{equation}

\subsection{The integrator}
\label{ssec:theinteg}

Once we have the explicit maps $\exp\left(\tau L_A\right)$ and $\exp\left(\tau
  L_B\right)$ one can construct the basic second order symplectic scheme
St\"ormer-Verlet/Leap Frog: $Y_{2}(\tau)  = e^{\frac{\tau}{2} L_A}e^{\tau
  L_B}e^{\frac{\tau}{2} L_A}$. Then Yoshida proved~\cite{Yo1990} that one can
find explicit suitable coefficients $x_1=\frac{1}{2-2^{1/3}}$ and $x_0=-2^{1/3}x_1$, such that the composition
\begin{equation}
\label{eq:Yo4}
 Y_{4}(\tau)  =  Y_{2}\left(x_1\tau \right)\circ Y_{2}\left(x_0\tau
 \right)\circ Y_{2}\left(x_1\tau \right)\, ,
\end{equation}
is actually a fourth order symplectic integrator, that moreover is symmetric,
i.e. it has exact time reversibility.

One can iterate this construction and find new explicit coefficients, $y_1=\frac{1}{2-2^{1/5}}$ and $y_0=-2^{1/5}y_1$, such that the composition
\begin{equation}
\label{eq:Yo6}
 Y_{6}(\tau)  =  Y_{4}\left(y_1\tau \right)\circ Y_{4}\left(y_0\tau
 \right)\circ Y_{4}\left(y_1\tau \right)\, ,
\end{equation}
provides a sixth order symmetric symplectic integrator. This idea can be
iterated to construct symmetric symplectic integrators with arbitrarily high
order. The main drawback is that the number of involved terms increases very
fast, thus one has to choose a suitable compromise between the required
precision in term of preserved quantities, i.e. energy and mass, and the CPU
time available.

In the next sections we will show that $Y_4$ and $Y_6$ exhibit very good
energy and mass preservation properties even for relatively large integration
time 
steps, they are composed by relatively few terms and 
moreover because they are explicit methods, they are relatively fast. They
provide thus very powerful and fast methods to numerically analyze the DNLS in
the asymptotic limit of large time spans.

\begin{remark}
  Let us observe that our method straightforwardly applies to generalized
  DNLS~\cite{FKS2009}
\begin{equation*}
  H_{gDNLS}(\vec{p},\vec{q})=-i\sum_{l=1}^N\left[\frac{(p_{l+1}-p_l)(q_{l+1}-q_l)}{h^2}-\left(p_lq_l\right)^{\sigma+2}\right]\,
  , 
\end{equation*}
being $\sigma$ a positive parameter. In fact setting
$A_{\sigma}(\vec{p},\vec{q})=i\sum_{l=1}^N(p_lq_l)^{\sigma+2}$ 
and observing that $C_l=p_lq_l$ is still a first integral for the flow of
$A_{\sigma}$, we directly obtain for $\exp(\tau L_{A_{\sigma}})$
\begin{equation*}
  \forall l\in\{1,\dots,N\}\quad 
  \begin{cases}
    p_l(\tau)&=e^{-i(\sigma+2)C^{\sigma+1}_l\tau}p_l(0)\\
    q_l(\tau)&=e^{i(\sigma+2)C^{\sigma+1}_l\tau}q_l(0)\, .
  \end{cases}
\end{equation*}
While the flow associated to the $B$ part remains unchanged.

More generally our method can be directly applied whenever we replace the
$A$--part of 
the Hamiltonian function with a new one, $A^{\prime}(\vec{p},\vec{q})$, for
which the map $\exp(\tau L_{A^{\prime}})$ can be computed explicitely.
\end{remark}

\subsection{Preserved quantities}
\label{ssec:presquant}

By construction the symplectic scheme $Y_{2m}\left(\tau\right)$ will preserve
the energy of the systems with an error $\mathcal{O}\left(\tau^{2m}\right)$,
that is independent of the 
relative weight of the functions $A$ and $B$ used to decompose the Hamilton
function, this is the reason why this method is more suitable than the one
proposed 
in~\cite{LR2001} where the error is also a function of the relative weights.

The aim of this section, is to prove that the symplectic schemes $Y_{2m}$
preserve other relevant quantities of the DNLS system~\eqref{eq:maineq}.

\subsubsection{Preservation of the first integral $I(\vec{p},\vec{q})=\sum_{l}p_lq_l$.}
\label{ssec:ifirst}
To prove that our method preserves the first integral $I$ let us start by
consider the action of the map $\exp\left(\tau L_B\right)$ on the function 
$I=\sum_lp_lq_l$. Starting by its very first definition~\eqref{eq:expB1} we get:
\begin{eqnarray*}
  \frac{d}{dt}I\rvert_{\text{flow $B$}}&=&-\frac{i}{h^2}\sum_{l=1}^N
  \left[\left(p_{l+1}-2p_l+p_{l-1}\right)q_l-p_l\left(q_{l+1}-2q_l+q_{l-1}\right)\right]
  \\
&=&-\frac{i}{h^2}\sum_{l=1}^N
  \left(p_{l+1}q_l+p_{l-1}q_l-p_lq_{l+1}-p_lq_{l-1}\right)=0\, ,
\end{eqnarray*}
where the last equality follows by using the boundary conditions. Thus the
flow of $B$ preserves $I$.

On the other hand by the definition of the map $\exp\left(L_A\right)$ given
by~\eqref{eq:expA3} one straightforwardly get:
\begin{equation*}
  p_l(t)q_l(t)=p_l(0)q_l(0) \quad \forall l\in\{1,\dots,N\}\, ,
\end{equation*}
and thus also the flow induced by $A$ preserves $I$.

We can thus conclude that the composition of the maps $\exp\left(L_A\right)$
and $\exp\left(L_B\right)$ preserves the first integral $I$ and so does any
symplectic scheme $Y_{2m}$.

\subsubsection{Preservation of the relation $\overline{\vec{p}(t)}=\vec{q}(t)$}
\label{ssec:presconj}

Let us define the complex vector
$\Delta(t)=\overline{\vec{p}(t)}-\vec{q}(t)$. Using the fact that $M$ is real,
the time evolution of $\Delta$ under the action of $B$ is given
by~\eqref{eq:expB2}: 
\begin{equation*}
  \frac{d}{dt}\Delta\rvert_{\text{flow
      $B$}}=-\frac{i}{h^2}M\overline{\vec{p}}+\frac{i}{h^2}M\vec{q}=-\frac{i}{h^2}M\Delta\, ,
\end{equation*}
from which one gets:
\begin{equation*}
  \Delta(t)=e^{-\frac{i}{h^2}Mt}\Delta(0)\, .
\end{equation*}
Thus if by assumption $\Delta(0)=0$ then we get $\Delta(t)=0$ for all $t$.

For the flow of $A$ we use once again the explicit map~\eqref{eq:expA3} and
the fact that $\overline{\vec{p}(0)}=\vec{q}(0)$ implies that
$C_l=p_l(0)q_l(0)=|q_l(0)|^2$ is real for all $l\in\{1,\dots,N\}$, thus
\begin{equation*}
  \overline{p_l(t)}=e^{2i\overline{C}_lt}\overline{p_l(0)}=e^{2iC_lt}q_l(0)=q_l(t)\, .
\end{equation*}

Once again, we can thus conclude that the composition of the maps
$\exp\left(L_A\right)$ and $\exp\left(L_B\right)$ preserves the first integral
$I$ and so it does any symplectic scheme $Y_{2m}$.

\subsubsection{Other preserved quantities}
\label{ssec:presnormpq}

The symplectic integrators we proposed preserve other quantities such as the
norm of the canonical variables, namely $|\vec{p}|^2$ and $|\vec{q}|^2$,
defined by the complex scalar product $|\vec{p}|^2=<\vec{p},\vec{p}>$.

Let us first observe that this statement holds for the flow of $B$. For
instance in the case of the $\vec{p}$ variable one has:  
\begin{eqnarray*}
\label{eq:presBp}
  \frac{d}{dt}<\vec{p},\vec{p}>&=&<\dot{\vec{p}},\vec{p}>+<\vec{p},\dot{\vec{p}}>=_{\text{Eq.}~\eqref{eq:expB2}}=\frac{i}{h^2}<M\vec{p},\vec{p}>-\frac{i}{h^2}<\vec{p},M\vec{p}>\\
&=&\frac{i}{h^2}<M\vec{p},\vec{p}>-\frac{i}{h^2}<M^{\dag}\vec{p},\vec{p}>=0\, ,
\end{eqnarray*}
where in the last step we used the facts that $M$ is real, thus
$M^{\dag}=M^{\rm T}$, and moreover $M^{\rm T}=M$.

It remains to check the behavior under $A$. But using the
definition~\eqref{eq:expA3} we get: 
\begin{equation*}
  |\vec{p}(t)|^2=<\vec{p}(t),\vec{p}(t)>=\sum_l\overline{p}_l(t)p_l(t)=\sum_le^{2iC_lt}\overline{p}_l(0)e^{-2iC_lt}p_l(0)=\sum_l\overline{p}_l(0)p_l(0)=|\vec{p}(0)|^2\, .
\end{equation*}

Along a very similar way we can prove the result for $\vec{q}$. We can thus
conclude that the symplectic integrators $Y_{2m}$ preserve the norm of the
complex vectors $\vec{p}$ and $\vec{q}$.

\section{Results}
\label{sec:results}

The aim of this section is to present the numerical results obtained using our
high order symplectic schemes. We fixed initial conditions as
in~\cite{Schober1999,KS2001} to have a good testbed to compare our results
with other ones available in the literature, hence we define:
\begin{equation}
  \label{eq:inicond}
  q_l(0)=\overline{p_l(0)}=a\left(1-\epsilon \cos (bx_l)\right)\, ,
\end{equation}
where $x_l=-L/2+(l-1)h$, $h=L/N$, $l\in\{1,\dots,N\}$, namely we are
considering perturbation of a spatially uniform plane wave invariant under the
phase flows. We used several values
for $N$ while the remaining parameters have
been fixed~\cite{Schober1999,KS2001} to $\epsilon=10^{-2}$, $b=2\pi/L$ and
$L=2\sqrt{2}\pi$. 

Let us stress that because of the high accuracy of most of the presented
results, mainly in term of energy preservation, we performed our numerical
simulation using quadruple precision Fortran.

\begin{figure}[htbp]
\begin{center}
\includegraphics[width=11.5cm]{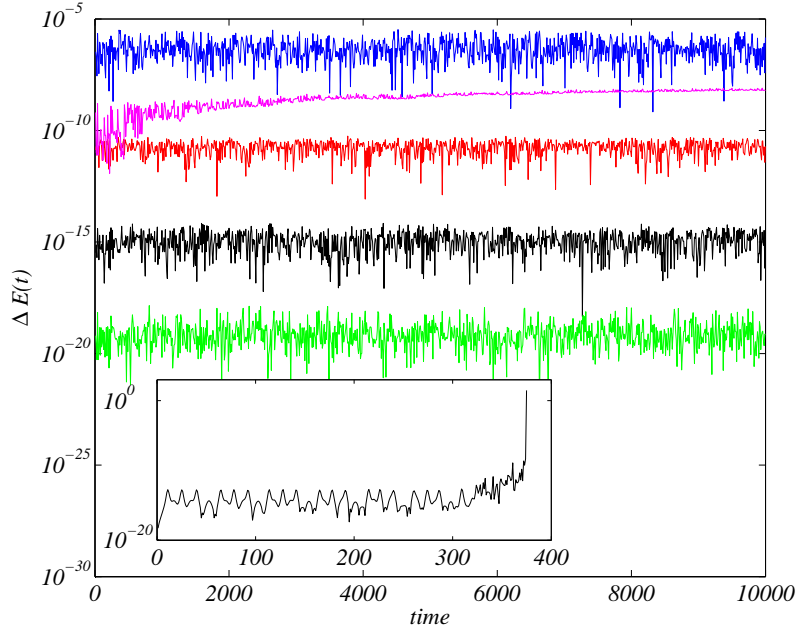}
\end{center}
\caption{{\em Time evolution of relative energy}. Semilog plot of the relative
  energy  
  loss, $\Delta E(t)=|E(t)-E(0)|/|E(0)|$, as a function of time, for one orbit
  with initial conditions~\eqref{eq:inicond}, $N=4$ and a 
  time step $\tau=0.01$. Using, from top to bottom, the integrators $Y_2$
  (blue), $RK_4^{mod}$ (magenta), $Y_4$ (red), $Y_6$ (black) and $Y_8$
  (green). Inset : using 
  $RK_4$ with time step $\tau=0.001$.}
\label{fig:nrgevol}
\end{figure}

The first result reported in Fig.~\ref{fig:nrgevol} shows the preservation of
the energy for one given orbit with the above initial conditions using the
symplectic 
schemes $Y_{2m}$ for $m\in\{1,2,3,4\}$. We can observe that the
relative energy loss, $\Delta 
E(t)=|E(t)-E(0)|/|E(0)|$, fluctuates in times but doesn't grow on a quite
large time span $[0,10^4]$ even using a relatively large time step
$\tau=10^{-2}$. Moreover we can remark that 
already with $Y_4$ the relative energy loss, is well below $10^{-10}$. On the
other hand $Y_8$ allows to reach values of the order of $10^{-20}$.

The preservation of the other quantities is even better; concerning the mass,
$I\left(\vec{p},\vec{q}\right)$, we can find that the relative error, $\Delta 
I(t)=|I(t)-I(0)|/|I(0)|$, assumes values well below $10^{-26}$ for all
integrator schemes we used $Y_{2m}$ (data not reported); while the relation
$\overline{\vec{p}(t)}=\vec{q}(t)$ is preserved up to the machine precision,
namely $\sim 10^{-34}$ (data not reported).  

 In the
inset of Fig.~\ref{fig:nrgevol} we report the comparison with the
non--symplectic integrator Runge--Kutta 4 using $\tau=10^{-3}$; we can clearly
see the inefficiency of the latter method even using a time step smaller than
the one used for the symplectic schemes, in fact the relative energy loss 
becomes larger than $10^2$ already at $t\sim 350$, so no longer comparison are
possible. Using Runge--Kutta 4 the relative energy loss grows in times, the
slower is the time step, but still growing, hence one can reach the precisions
obtained by a symplectic integrator only using very small time steps $\tau$ or
integrating over relatively short time spans. For instance one can achieve a
relative energy loss of $\sim 10^{-5}$ using $RK_4(10^{-4})$ only on a time span
$[0,\sim 350]$, while using $Y_2(10^{-2})$ we can get the same error but on
$[0,10^4]$. On the time span $[0,\sim 350]$ and using the time step
$\tau=10^{-4}$, Runge--Kutta 4 achieves a relative loss for the mass of the
order of $10^{-8}$ while the conjugacy relation 
$\overline{\vec{p}(t)}=\vec{q}(t)$ is not preserved any more. 

This is the main
reason of the poor properties obtained using Runge--Kutta 4 with $\tau=0.01$;
in fact if we modify the integration scheme by forcing the vectors
$\vec{p}(t)$ and $\vec{q}(t)$ to satisfy the conjugation relation at each time
step, we obtain a method, hereby called modified Runge--Kutta 4,
$RK_4^{mod}$, that exhibits improved energy preservation properties (see
Fig.~\ref{fig:nrgevol}). Let us observe that this is method is not symplectic,
as one can clearly conclude from the increasing trend in the relative energy
variation presented in the Figure. The method allows to reach large time
spans, $[0,10^4]$, but the goodness of the method, measured in terms of
relative energy variation, is worse than $Y_4$ and slightly better than
$Y_2$. Let us finally observe that on the same large time span and still using
$\tau=0.01$, the mass of the system is conserved up to a factor $1.5\, 10^{-8}$.

\begin{figure}[htbp]
\begin{center}
\includegraphics[width=11.5cm]{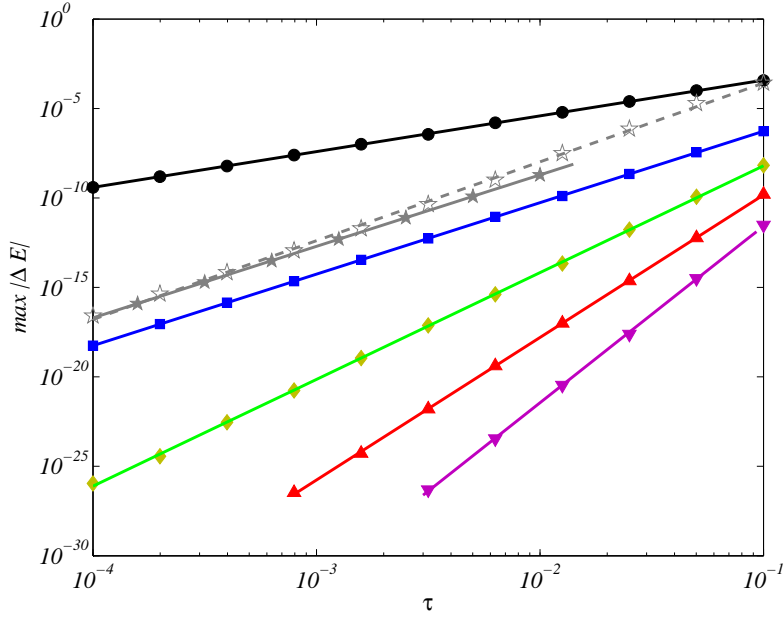}
\end{center}
\caption{{\em Accuracy of the integrators as a function of
    $\tau$}. We report
  $\max_{t\in[0,10^4]}|\Delta E(t)|$ for orbits with initial conditions given
  by~\eqref{eq:inicond} and $N=4$, numerically integrated over the time span
  of $[0,10^4]$ with $\bigcirc$ 
  $Y_2$ (black), $\square$ $Y_4$ (blue), $\Diamond$ $Y_6$ (green), $\bigtriangleup$
  $Y_8$ (red), $\bigtriangledown$ $Y_{10}$ (cyan), empty $\star$
  $RK_4^{mod}$ (grey) and $\star$ $RK_4$ (grey) but over a small time span
  $[0,200]$. Straight lines are
  linear best fits $\log_{10}\max|\Delta E|=\alpha \log_{10}\tau+\beta$, the
  coefficients $\alpha$ and $\beta$ are reported in Table~\ref{tab:lifit}.}
\label{fig:nrgloss}
\end{figure}

The next study will concern the dependence of the maximum relative energy
loss as a function of the time step, by integrating several solutions using
$Y_{2m}\left(\tau\right)$, $m\in\{1,2,3,4,5\}$, over a large time span
$[0,10^4]$; let us observe, 
once again, that because the relative energy fluctuates around the fixed
initial value, the same results hold for arbitrarily longer time spans. Result
reported in Fig.~\ref{fig:nrgloss} shows the computed numerical accuracy of
$Y_{2m}\left(\tau\right)$ as a function of the time step $\tau$. Let us
observe that the use of $Y_{10}$, respectively of $Y_{8}$, with integration steps
$\tau$ smaller than $\sim 2.\, 10^{-3}$, respectively $\sim 8.\,10^{-4}$, will
produce a maximum relative energy loss below the quadruple machine
precision, that is why we limited our simulations to these values. Straight
lines 
reported in Fig.~\ref{fig:nrgloss} represent linear best fits
$\log_{10}\max|\Delta E|=\alpha \log_{10}\tau+\beta$, whose coefficients
$\alpha$ and $\beta$ are reported in Table~\ref{tab:lifit} and numerically
confirm the accuracy of the integrators $Y_{2m}$. 

Let us remark that a similar result holds for larger
values of $N$. The main difference being that in this case large time steps
are prevented to be used because of a decrease in the performances of all
integrator schemes, $Y_{2m}(\tau)$, mainly because of the energy
preservation. This fact has 
been already observed~\cite{KS2001} and relies on a stability issue of the
integrators, symplectic and non--symplectic ones, that imposes a constrains
$\tau N \approx c$, for some positive constant $c$.

\begin{table}[h]
\begin{tabular}{|c|c|c|}
\hline
integrator & $\alpha$ & $\beta$\\
\hline\hline
$Y_2$ & $1.997\pm 0.012$ & $-1.42\pm 0.03$\\
$Y_4$ & $3.996\pm 0.009$ & $-2.271\pm 0.023$\\
$Y_6$ & $5.97\pm 0.06$   & $-2.23\pm 0.15$ \\
$Y_8$ & $7.97\pm 0.07$   & $-1.86\pm 0.16$ \\
$Y_{10}$& $9.88\pm 0.17$  & $-1.7\pm 0.3$\\
$RK_4$& $3.997\pm 0.001$  & $-0.716\pm 0.006$\\
$RK_4^{mod}$& $4.40\pm 0.11$  & $0.82\pm 0.29$\\
\hline 
\end{tabular}
\caption{Numerical coefficients of the linear best fits $\log_{10}\max|\Delta
  E|=\alpha \log_{10}\tau+\beta$ reported in Fig.~\ref{fig:nrgloss}.}
\label{tab:lifit}
\end{table}

\begin{figure}[htbp]
\begin{center}
\includegraphics[width=11.5cm]{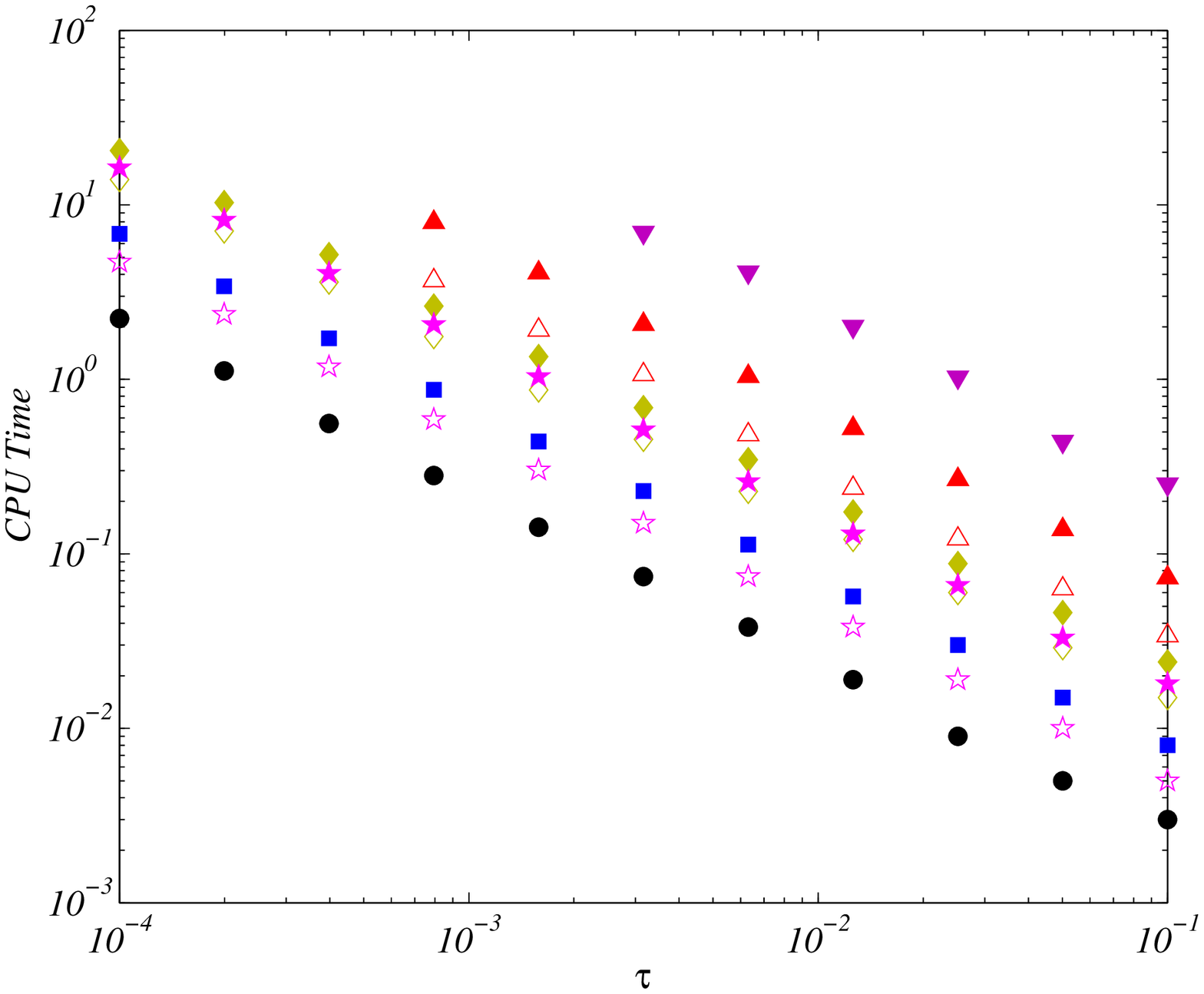}
\end{center}
\caption{{\em CPU Time needed by $Y_{2m}\left(\tau\right)$ as a function of
    $\tau$}. For a fixed time span of $[0,10^4]$ we report
  the CPU time needed by $Y_{2m}\left(\tau\right)$ to integrate orbits with
  initial conditions given by~\eqref{eq:inicond} and $N=4$. Symbols are :
  $\bigcirc$ 
  $Y_2$ (black), $\square$ $Y_4$ (blue), $\Diamond$ $Y_6$ (green), empty
  $\Diamond$ 
  $Y^{opt}_6$ (green), $\bigtriangleup$
  $Y_8$ (red), empty $\bigtriangleup$
  $Y^{opt}_8$ (red), $\bigtriangledown$ $Y_{10}$ (cyan), empty $\star$
  $RK_4^{mod}$ (magenta) and $\star$ $RK_4$ using (magenta).}
\label{fig:cputime}
\end{figure}

Our last remark concerns the speed of the numerical methods $Y_{2m}(\tau)$. In
Fig.~\ref{fig:cputime} we report the CPU time used to integrate orbits with
initial conditions~\eqref{eq:inicond} and $N=4$ using $Y_{2m}(\tau)$, as a
function of the time step $\tau$. For $Y_8$ and $Y_{10}$ we limited the
analysis to time steps whose maximum relative energy loss is larger than the
machine precision using quadruple precision Fortran (see
Fig.~\ref{fig:nrgloss} and discussion therein). From these data we clearly
conclude that the CPU time increases
as $1/\tau$ for a fixed integrator scheme $Y_{2m}(\tau)$; on the other hand,
for fixed $\tau$, the CPU time increases as a exponential of the integrator
order, roughly as $3^m$. This is because, as already mentioned, the Yoshida
scheme 
does not get the optimal number of products $\exp(L_A)$ and $\exp(L_B)$. On one
hand Yoshida already proposed~\cite{Yo1990} an improved version to tackle this
difficulty, whose results are symplectic schemes with fewer
compositions~\eqref{eq:approxflow}. Because the computations to obtain the
coefficients $(c_j,d_j)$ become rapidly cumbersome we limit ourselves to
study the cases of sixth order, $Y^{opt}_6$, and eighth order,
$Y^{opt}_8$. Results reported in Fig.~\ref{fig:cputime} show that $Y_6(\tau)$
needs almost $1.5$ times more CPU time than $Y^{opt}_6(\tau)$, while $Y_8(\tau)$
about $2.2$ times more CPU time than $Y^{opt}_8(\tau)$.

On the other hand in
practical applications one should choose the time step $\tau$ and the
integration order $2m$ that produce the best balance between the required
precision, say maximum relative energy loss, and the CPU time needed. For
instance from
Fig.~\ref{fig:nrgloss} we clearly see that using $Y_4$ and $\tau\sim 10^{-3}$,
we can ensure a maximum relative energy loss of the order 
of $10^{-15}$, to get the same precision using $Y_6$ one can use a time step
ten times larger, $\tau\sim 10^{-2}$ and thus (see Fig.~\ref{fig:cputime}) the
required CPU time will be smaller 
with $Y_6(10^{-2})$ than with $Y_4(10^{-3})$. Requiring the same precision,
using instead $Y_8$, will need a time step \lq\lq only\rq\rq twenty times
larger and thus the CPU time will increase: $Y_8(2.\,10^{-2})>Y_6(10^{-2})$.

The Runge--Kutta 4 requires a CPU time larger than $Y_4$ using the same time
step, roughly of the same order of $Y_6^{opt}$. On the other hand the modified
Runge--Kutta scheme is faster than $Y_4$ because it has to compute only half
of the vector field.

The dependence of the CPU time on the discretization parameter $h$, hence on
$N$, is more crucial once we need to use very large $N$ and/or a very large
number of orbits. Because in the present work we were not interested in the
optimality of the method, we computed \lq\lq naively\rq\rq the map
$\exp(L_B)$, namely using a vector--matrix product whose cost is
$\mathcal{O}\left(N^2\right)$. On the other hand we can 
easily improve this part by considering the strong similarity with the map
$\exp(L_B)$ and the Fourier transform of the vectors
$\left(\vec{p},\vec{q}\right)$ (see 
Proposition~\ref{prop:circmatr} and Eq.~\eqref{eq:expB5}) and thus use,
instead of the vector--matrix product, a Fast Fourier--like method to speed up
the computations. 

\section{Conclusions}
\label{sec:concl}

In this paper we presented a family of high order, explicit, symplectic
integrations schemes adapted to the study of the DNLS. Despite DNLS has been
studied numerically since long time, this is the first time that such a high
precision can be achieved using relatively large time steps. Besides the very
good energy 
preservation properties of the above introduced methods, we also obtained an
almost exact preservation of the other first integral, the mass of the system,
and of the conjugacy relation. Because the integrators we constructed are
explicit ones, they result very fast.

For all these reasons we believe that such accurate numerical schemes could be
very useful to test several physical hypotheses concerning the asymptotic
regimes of the DNLS, for instance the existence and stability of breathers and
the regimes with negative temperature.

\section*{Acknowledgments}

One of the authors, TC, would like to thank Antonio Politi and Stefano Iubini,
from ISC Florence Italy, for interesting and useful discussions. The work of
Ch.H is supported by a FNRS Research Fellowship.  Numerical simulations were
made on the local computing resources (Cluster URBM-SYSDYN) at the University
of Namur (FUNDP, Belgium).

\end{document}